\documentclass[a4paper]{llncs}

\usepackage{amsmath}
\usepackage{graphicx}
\usepackage{xcolor}
\usepackage{hyperref}
\usepackage{algorithm}
\usepackage{listings}
\usepackage{tabularx}
\usepackage{times}
\usepackage{amssymb}
\usepackage{comment}

\lstdefinelanguage{GOTO}{
    morekeywords = [1]{IF, GOTO, THEN},
    morecomment = [l]{//}
}

\usepackage[textsize=footnotesize]{todonotes}

\newcommand\KK[1]{}
\newcommand\GR[1]{}
\newcommand\FB[1]{}
\newcommand\Fedor[1]{}
\newcommand\XL[1]{}

\newcommand{\changed}[1]{#1}
\newcommand{\Changed}[1]{#1}

\title{ESBMC v7.4: Harnessing the Power of Intervals}
\subtitle{(Competition Contribution)}

\author{Rafael Menezes\inst1
    \and Mohannad Aldughaim\inst1\inst5
    \and Bruno Farias\inst1
    \and Xianzhiyu Li\inst1
    \and Edoardo Manino\inst1
    \and Fedor Shmarov\inst1\inst4
    \and Kunjian Song\inst1
    \and Franz Brau\ss{}e\inst1
    \and Mikhail R. Gadelha\inst2
    \and Norbert Tihanyi\inst3
    \and Konstantin Korovin\inst1
    \and Lucas C. Cordeiro\inst1}
\institute{
  University of Manchester, UK \and Igalia, A Coruña, Spain
  \and Eötvös Loránd University, Hungary
  \and Newcastle University, UK
  \and King Saud University, Saudi Arabia
}

\begin{document}
\maketitle


\begin{abstract}
ESBMC implements many state-of-the-art techniques for model checking. We report
on new and improved features that allow us to obtain verification results for
previously unsupported programs and properties. ESBMC employs a new static
interval analysis of expressions in programs to increase verification
performance. This includes interval-based reasoning over booleans and integers,
forward and backward contractors, and particular optimizations related to
singleton intervals because of their ubiquity.
Other relevant improvements concern the verification of concurrent
programs, 
as well as several operational models, internal ones, and also those of
libraries such as pthread and the C mathematics library. 
An extended memory safety analysis now allows tracking of memory leaks that are
considered still reachable.
\end{abstract}


\section{Verification Approach} 


ESBMC~\cite{CordeiroFM12,GadelhaMMC0N18} is a context-bounded model checker for the verification of single- and multi-threaded C programs for various code safety violations  (e.g., buffer overflows, dangling pointers, arithmetic overflows)  and user-defined assertions. It has been successfully participating in the SV-COMP competitions for many years due to our continuous work towards improving its performance. ESBMC transforms a given C program using a Clang-based~\cite{LLVM:CGO04} front-end into an intermediate representation in the GOTO language~\cite{10.1007/978-3-540-24730-2_15}, which is 
symbolically executed to produce verification formulae passed to one or more SMT solvers. 



\section{Software Architecture\FB{Can we rename this section to reflect its contents? Sth.\ like `Improvements over v7.0' or similar?}}
\paragraph{Interval analysis}

In this year, ESBMC interval analysis was improved using
Abstract Interpretation techniques~\cite{cousot2021principles}.
\FB{Did we really decide this for SV-COMP? \Changed{Yes, the wrapped domain (should we describe it here?) was also an option}}%
We used the integer domain (with infinities) as the abstract domain for SV-COMP. The domain consists of, for each variable in the program, keeping the box interval (i.e., a \emph{minimum} and \emph{maximum}) for all statements.
ESBMC also supports interval arithmetic and widening strategies (through extrapolation and interpolation). Once computed, the intervals are used for optimizations and code instrumentation.

\pagebreak[2]

Two optimizations are used: \emph{singleton propagation} and \emph{guard elimination}\footnote{ESBMC refers to path conditions as guards.}. The singleton optimization 
applies when a variable used in an expression is known to have only one possible value; therefore we
replace the variable \changed{in the expression} with the interval value. For guard elimination, we consider
\FB{Does `cast of interval into' mean evaluating a comparison of the interval to zero / some scalar?}%
a cast of the interval into
\FB{This is \href{https://en.wikipedia.org/wiki/Three-valued_logic\#Kleene_and_Priest_logics}{Kleene's} three-valued logic $K_3$, I suppose. Should we mention/cite it? \Changed{Sure... I am not sure from which logic we are based on.}}%
a three-valued boolean domain (i.e., \emph{true}, \emph{false}, \emph{maybe}). Using this abstract domain, we can optimize always true or false conditions. This is applied recursively in the expression (i.e., try the full expression and then on its operands), being able to reason on both operands and full expression. After the optimization, 
ESBMC runs a
constant propagation pass over the new expression.

Regarding the new code instrumentation, the main use of intervals is to generate invariants 
which the $k$-induction strategy benefits from most.
This is done by adding assumptions 
restricting the value of variables, e.g., $x > 3 \wedge x < 10$.
In previous editions, ESBMC would instrument an assumption at the start of a loop (before and inside) with all function variables (even if they do not affect the loop variables). For this year, we changed the approach to instrument the assumption only for
\changed{variables that are part of the statement,}
e.g.,
for an \emph{``if"} statement, we only add an assumption with the variables in the condition and the path condition.
Lastly, we expanded the types of instrumented statements: assertions, conditionals, and function calls.

\paragraph{Contractors}
\changed{%
A contractor $\mathcal{C}:\mathbb{R}^n\to\mathbb{R}^n$ 
is an interval method that approximates the solutions of a Constraint Satisfaction Problem (CSP)~\cite{MUSTAFA2018160} over a set of variables $\mathrm{x}$ and variable domains $[\mathrm{x}] \in \mathbb{IR}^n$ with the following properties.
}
For 
solution set $\mathbb{S}_{\mathrm{x}}$ and a contractor $\mathcal{C}$, when applied to a box $[\mathrm{x}]$, it ensures that $\mathcal{C} ([\mathrm{x}])\subseteq [\mathrm{x}]$ and $\mathcal{C}([\mathrm{x}])\cap\mathbb{S}_{\mathrm{x}}=
[\mathrm{x}]\cap\mathbb{S}_{\mathrm{x}}$, meeting both contraction and correctness conditions respectively~\cite{Jaulin2001AppliedIA}.

In ESBMC,  we apply 
contractors on conditional statements (\emph{``if''} and loops) by using their conditions to prune the search space for the variables domains.
This is done by partitioning the domain $[\mathrm x]$ into the three sets $\mathbb{S}_{\textit{in}}$, $\mathbb{S}_{\textit{boundary}}$ and $\mathbb{S}_{\textit{out}}$ such that
$\mathbb{S}_{\textit{in}}\subseteq\mathbb S_{\mathrm{x}}$ is an under-approximation of the solution set and $\mathbb{S}_{\textit{out}}\subseteq[\mathrm x]\setminus\mathbb S_{\mathrm{x}}$ is an under-approximation of the non-solutions.
These sets are then used to refine the ranges of each variable occurring in the conditional statement.

The Forward-backward Contractor is utilized 
\changed{by ESBMC}
for its simplicity and effectiveness. This type of contractor is specifically designed for CSPs with a single constraint~\cite{Granvilliers1999RevisingConsistency,hansen2003global,neumaier1990interval}. It operates in two stages: forward evaluation and backward propagation~\cite{MUSTAFA2018160,10.1007/978-3-031-30826-0_18}. In scenarios with multiple constraints in a CSP, the forward-backward contractor is applied to each constraint independently. 
\changed{%
ESBMC uses the library Ibex~\cite{ibex} to implement contractor-based reasoning
.
%
Ibex is implemented in C++ 
and designed for accurate interval arithmetic and constraint processing}.

\paragraph{Memory leaks}

This year, ESBMC employs a refined check for the \emph{valid-memtrack} property. This property is loosely described as only allowing those dynamically allocated objects to survive that are still reachable at the end of the program's execution by following a path of pointers stored in objects eventually referenced by global variables.
A property violation witness has to contain proof of \emph{unreachability} of a dynamic allocation starting from any global variable.

The new algorithm leverages the existing one tracking the lifetime of allocations for the \emph{valid-memcleanup} property, but it specifically excludes still-reachable objects from the check. This condition is encoded into an SMT formula using the paths deterministically described by expressions of type struct, union, pointer, or array with constant size.
Each possible successor along the path is obtained through the value-set, and the validity is encoded through guards 
which have to hold at the end of execution.

\paragraph{C mathematical library}

ESBMC v7.4 offers extended support for the \texttt{math.h} library. Accurate modeling of the semantics of this library is crucial for reasoning on the behavior of complex floating-point software. For example, most neural network code relies on $32$-bit floats and may invoke the \texttt{math.h} library to compute the result of activation functions, positional encodings, and vector normalisations~\cite{manino2023neurocodebench}.

In this respect, the IEEE 754 standard~\cite{IEEE754} mandates bit-precise semantics for a small subset of the \texttt{math.h} library only. 
This subset includes addition, multiplication, division, \texttt{sqrt}, \texttt{fma}, and other support functions such as \texttt{remquo}. In contrast, the behavior of most transcendental functions (e.g., \texttt{sin}, \texttt{cos}, \texttt{exp}, \texttt{log}) is platform-specific. Still, the standard recommends implementing the correct rounding whenever possible.

As a tradeoff between precision and verification speed, ESBMC now features a two-pronged design. For the most commonly-used \texttt{float} functions, we borrow the MUSL plain-C implementation of numerical algorithms~\cite{MUSL}. For the corresponding \texttt{double} functions, we employ less complex algorithms with approximate behavior.


\paragraph{Data races}
Data races occur when multiple threads concurrently access the same memory location, and at least one of these accesses involves a write operation.
ESBMC's algorithm for checking data races 
\changed{%
extends
the static code instrumentation
CBMC~\cite{10.1007/978-3-540-24730-2_15}
uses.
The idea is 
to add a 
flag $A'$ to each variable $A$ involved in 
an assignment.
For instance, when $A$ is assigned a value, we create a new variable 
$A'$ 
and set it to true.
Directly after the assignment to $A$, 
we reset $A'$ to false.}
To identify races, we assert that the value of $A'$ is false when $A$ is accessed. Subsequently, we outline the challenges encountered by ESBMC and the improvements we have implemented.

\changed{%
As this method introduces additional instructions into the program, the potentially larger number of thread interleavings is counteracted by inserting atomic blocks appropriately.}
\changed{%
Data races are now also checked on access of arrays with non-constant indices.}
\changed{%
The most challenging aspect of data race detection is the dereference of pointers, as the pointee would have to be instrumented but is not statically known through the value-set analysis.}
Thus, 
\changed{the} new implementation is 
\changed{hybrid}, addressing cases unsuitable for static analysis during symbolic execution, 
\changed{thereby enabling ESBMC to detect} more types of data races.


\section{Strengths and Weaknesses}

\paragraph{Interval analysis} The interval analysis improved and provided better invariants for ESBMC. The new optimizations help ESBMC to solve new benchmarks in categories with multiple path conditions (i.e., ECA).
\Changed{The main weakness of the method is that our Abstract Interpreter only has partial support for widening, and it is not context-aware (i.e., function parameters and global variables cannot be tracked globally)}.
This results in a slowdown for categories with \Changed{loops with thousands of statements} (e.g., Hardware). For this reason, we had to go for more imprecise intervals by disabling interval arithmetic (arithmetic operations are extrapolated to infinity).

\paragraph{Contractors} While contractors are highly regarded for their ability to provide assured limits on solutions, their cautious approach may \changed{lead to} overly broad results and less precise conclusions. Therefore, a more rigorous evaluation of contractors is essential to assess their advantages and limitations effectively.

\paragraph{Data races}
    From the results, the data race detection of ESBMC v7.4 is promising. Compared 
    to the 
    previous version, the new algorithm 
    supports more types of 
    expressions and reduces the verification time.
    \changed{%
    The relatively high number of 2.2\% incorrect-true verdicts is mostly due to missing support for detecting data races during dereferences of pointers to compound types.
    This issue will addressed in the future work.}

\paragraph{C mathematical library}
Without operational models of the \texttt{math.h} library,
ESBMC would assign non-deterministic results, which may cause incorrect counterexamples to be returned. This behavior is especially evident for older versions of ESBMC on neural network code~\cite{manino2023neurocodebench}, as it usually contains many mathematical operations. ESBMC v7.4 fixes this semantic issue by providing explicit operational models for many common functions in \texttt{math.h}, thus yielding no incorrect results on the benchmarks in~\cite{manino2023neurocodebench}, \Changed{and achieving second place in the ReachSafety-Floats sub-category.}

\paragraph{Memory leaks}
    The new algorithm for the \emph{valid-memtrack} sub-property allowed ESBMC to identify 70 
    / 153 violations correctly with
    no incorrect verdicts
    (previous year: 0 / 134).
    There is a theoretical weakness in the current implementation concerning dynamic allocations only reachable through pointers stored in arrays of statically unknown size. It 
    could result in spurious incorrect-false verdicts but has not been 
    observed in test cases, yet.
    We will address this weakness and submit suitable tasks for this property to SV-COMP in the future.



\section{Tool Setup and Configuration}


ESBMC can be used via the python wrapper \texttt{esbmc-wrapper.py} to simplify its usage for the competition. Please refer to its help message (\texttt{-h}) for usage instructions.
\FB{Should we mention sth.\ like `and selects the solving strategy based on features of the benchmark, such as usage of pthread or keras2c functions'?}%
This wrapper runs the ESBMC executable with command line options specific to each supported property.

\section{Software Project}


ESBMC is a joint project with the Federal University of Amazonas (Brazil),  University of Southampton (UK), University of Manchester (UK),  and University of Stellenbosch (South Africa). It is publicly available at
\changed{\url{http://esbmc.org} under the terms of the Apache License 2.0 and %
static release builds of ESBMC are provided at \url{https://github.com/esbmc/esbmc}}.
The exact version that participated in SV-COMP 2024 is available at \url{https://doi.org/10.5281/zenodo.10198805}.



\pagebreak[3]

\bibliographystyle{abbrv}
\bibliography{main}

\end{document}